\documentclass{article}
\usepackage{spconf}

\usepackage{amsmath,amssymb,amsfonts}
\usepackage{algorithmic}
\usepackage{graphicx}
\usepackage{textcomp}
\usepackage{xcolor}
\usepackage{tikz}
\usepackage{standalone}
\usepackage{flushend} 

\newcommand{\naive}{na\"ive }
\newcommand{\Naive}{Na\"ive }

\newcommand{\secnegspace}{\vspace{-0.11cm}}
\newcommand{\subsecnegspace}{\vspace{-0.17cm}}
\usepackage{ml_package}

\title{Calibrating~AI~Models for Few-Shot~Demodulation via~Conformal~Prediction}

\name{Kfir~M.~Cohen$^1$,\qquad Sangwoo~Park$^1$,\qquad Osvaldo~Simeone$^1$ \qquad Shlomo~Shamai~(Shitz)$^2$ 
        \thanks{The work of K. M. Cohen, S. Park and O. Simeone has been supported by the European Research Council (ERC) under the European Union’s Horizon 2020 research and innovation programme, grant agreement No. 725731. The work of O. Simeone has also been supported by an Open Fellowship of the EPSRC.}
        \thanks{The work of S. Shamai has been supported by the European Union's Horizon 2020 Research And Innovation Programme, grant agreement No. 694630.}
        \thanks{The authors acknowledge use of the research computing facility at King’s College London, Rosalind (https://rosalind.kcl.ac.uk).}}
\address{$^1$ KCLIP Lab, Department of Engineering, \emph{King’s College London}, UK.\\
$^2$ Viterbi Faculty of Electrical and Computing Engineering, \emph{The Technion}, Haifa, Israel.}

\begin{document}
%
\maketitle
\begin{abstract}
AI tools can be useful to address model deficits in the design of communication systems. However, conventional learning-based AI algorithms yield poorly calibrated decisions, unabling to quantify their outputs uncertainty. While Bayesian learning can enhance calibration by capturing epistemic uncertainty caused by limited data availability, formal calibration guarantees only hold under strong assumptions about the ground-truth, unknown, data generation mechanism. We propose to leverage the conformal prediction framework to obtain data-driven set predictions whose calibration properties hold irrespective of the data distribution. Specifically, we investigate the design of baseband demodulators in the presence of hard-to-model nonlinearities such as hardware imperfections, and propose set-based demodulators based on conformal prediction. Numerical results confirm the theoretical validity of the proposed demodulators, and bring insights into their average prediction set size efficiency.
\end{abstract}



\secnegspace
\section{Introduction} \label{sec: Introduction}

\begin{figure}
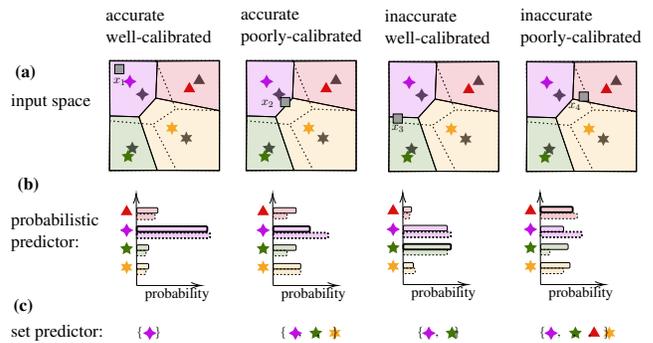

    \begin{minipage}[b]{1.0\linewidth}
        \centering
        \hspace{-0cm}
        \centerline{\includestandalone[width=8.5cm]{Figs/fig_tikz_qpsk_prob_pred}}
        \vspace{-0.4cm}
        \caption{QPSK demodulation with a demodulator trained using a limited number of pilots (gray symbols):  \textbf{(a)} Constellation symbols (colored markers), optimal hard prediction (dashed lines), and model trained using the few pilots (solid lines). Accuracy and calibration of the trained predictor depend on the test input (gray square). \textbf{(b)} Probabilistic predictors obtained from the trained model (solid bars) and optimal predictive probabilities (dashed bars), with thick line indicating the hard prediction. \textbf{(c)} Set predictors output a subset of the constellation symbols for each input.
        }
        \label{fig: fig_tikz_qpsk_prob_pred}
        \vspace{-0.5cm}
    \end{minipage}
\end{figure}

Artificial intelligence (AI) models typically report a confidence measure associated with each prediction, which reflects the model's \emph{self-evaluation} of the accuracy of a decision. Notably, neural networks implement \emph{probabilistic predictors} that produce a probability distribution across all possible values of the output variable. As an example, Fig.~\ref{fig: fig_tikz_qpsk_prob_pred} illustrates the operation of a neural network-based demodulator \cite{park2021fewpilots, kim2018communication, jiang2019mind}, which outputs a probability distribution on the constellation points given the corresponding received baseband sample. The self-reported model confidence, however, may not be a reliable measure of the true, unknown, accuracy of the prediction, in which case we say that the AI model is \emph{poorly calibrated}. Poor calibration may be a substantial problem when AI-based decisions are  processed within a larger system such as a communication network.

Deep learning models tend to produce either overconfident decisions when designed following a frequentist framework \cite{Guo2017Calibration}; or else calibration levels that rely on strong assumptions about the ground-truth, unknown, data generation mechanism when Bayesian learning is applied \cite{masegosa2020learning, morningstar2022pacm, zecchin2022robustpacm,cannon2022investigating,frazier2020model,ridgway2017probably}. This paper investigates the adoption of \emph{conformal prediction (CP)} \cite{vovk2005algorithmic, shafer2008tutorial, fontana2020conformal} as a framework to design provably well-calibrated AI predictors, with \emph{distribution-free} calibration guarantees that do not require making any assumption about the ground-truth data generation mechanism.

Consider again the example in Fig.~\ref{fig: fig_tikz_qpsk_prob_pred},  which corresponds to the problem of designing a demodulator for a QPSK constellation in the presence of an I/Q imbalance  that rotates and distorts the constellation. The hard decision regions of an optimal demodulator and of a data-driven demodulator trained on few pilots are displayed in panel (a), while the corresponding probabilistic predictions for some outputs are shown in panel (b). Depending on the input, the trained probabilistic model may result in either accurate or inaccurate hard predictions, whose accuracy is correctly or incorrectly characterized, resulting in well-calibrated or poorly calibrated 
predictions. Note that a well-calibrated probabilistic predictor should provide output probabilities close to the optimal predictor (dashed lines in panel (b)). Importantly, accuracy and calibration are distinct requirements.

CP leverages probabilistic predictors as a starting point to construct well-calibrated \emph{set predictors}. Instead of producing a probability vector (as in Fig.~\ref{fig: fig_tikz_qpsk_prob_pred}(b)), a set predictor outputs a subset of the output space (see Fig.~\ref{fig: fig_tikz_qpsk_prob_pred}(c)). A set predictor is well-calibrated if it contains the correct output with a pre-defined probability selected by the system designer. This paper introduces CP-based  demodulators, obtaining set predictors that satisfy formal calibration guarantees that hold irrespective of the ground-truth, unknown, 
distribution. The proposed approach is particularly relevant in practical situations characterized by a limited number of pilots, in which characterizing uncertainty is of critical importance.

In the rest of the paper, we first define the problem and present preliminaries in Sec.~\ref{sec: Problem Definition}. Then, we introduce CP-based set predictors in Sec.~\ref{sec: Conformal Prediction}, and describe experiments and conclusions in Sec.~\ref{sec: Experiments and Conclusions}.

\secnegspace
\section{Problem Definition}\label{sec: Problem Definition}

\subsecnegspace
\subsection{Channel Model}

Following the example in Fig.~\ref{fig: fig_tikz_qpsk_prob_pred}, we consider a communication link subject to phase fading and to unknown hardware distortions at the transmitter side \cite{park2021fewpilots, cohen2021learning}. Our goal is to design a well-calibrated data-driven set demodulator based on the observation of a few pilots. We follow the unconventional notation of denoting as $y[i]$ the $i$-th transmitted symbol, and as $x[i]$ the corresponding received sample. This will allow us to write expressions for the demodulator in a more familiar way, with $x$ representing the input and $y$ the output. Each frame consists of $N$ pilots symbols and data symbols. The pilots define a data set $\D=\{ z[i] \}_{i=1}^N$ of $N$ examples of input-output pairs $z[i]=(x[i],y[i])$ for $i=1,\ldots,N$, which is available to the receiver for the design of the demodulator.

Each transmitted symbol $y[i]$ is drawn uniformly at random from a given constellation $\mathcal{Y}$ \cite{demeng2018two}. For any given frame, the received sample $x[i]$ can be written as
\begin{equation}
    \rv{x}[i] = e^{\jmath \rvpsi} f_{\text{IQ}}(\rv{y}[i]) + \rv{v}[i], \label{eq: demodulation channel model}
\end{equation}
for a random phase $\rvpsi \sim \mathrm{U} [0,2\pi)$, where the additive noise is $ \rv{v}[i] \sim \mathcal{CN}(0,\SNR^{-1}) $ for a given signal-to-noise ratio level $\SNR$. Furthermore, the I/Q imbalance function \cite{tandur2007joint} is 
\begin{equation}
    f_{\text{IQ}}(\rv{y}[i]) 
    = \bar{\rv{y}}_{\text{I}}[i] + \jmath \bar{\rv{y}}_{\text{Q}}[i] , \label{eq:iq_imbalance}
\end{equation}
where
\begin{equation}
    \begin{bmatrix} 
        \bar{\rv{y}}_{\text{I}}[i] \\ \bar{\rv{y}}_{\text{Q}}[i]
    \end{bmatrix}
    =
    \begin{bmatrix} 
        1+\rvepsilon & 0 \\ 0 & 1-\rvepsilon
    \end{bmatrix}
    \begin{bmatrix} 
        \cos \rvdelta & -\sin \rvdelta \\ -\sin \rvdelta & \cos \rvdelta
    \end{bmatrix}
    \begin{bmatrix} 
        \rv{y}_{\text{I}}[i] \\ \rv{y}_{\text{Q}}[i]
    \end{bmatrix}, \label{eq: iq imbalance as matrix}
\end{equation}
with $\rv{y}_{\text{I}}[i] $ and $\rv{y}_{\text{Q}}[i]$ being the real and imaginary parts of the modulated symbol $\rv{y}[i]$; and  $\bar{\rv{y}}_{\text{I}}[i]$  and $\bar{\rv{y}}_{\text{Q}}[i]$ standing for the real and imaginary parts of the transmitted symbol $f_{\text{IQ}}(\rv{y}[i])$.  The channel parameters $\rvpsi$, $\rvepsilon$, and $\rvdelta$ are generated independently in each frame from a common distribution.

\subsecnegspace
\subsection{Probabilistic Predictors}\label{sec: Probabilistic Predictors}

\emph{Probabilistic predictors} implement a parametric conditional distribution model $p(y|x,\phi)$ on the output $y \in \mathcal{Y}$ given the input $x \in \mathcal{X}$, where $\phi\in\Phi$ is a vector of model parameters. Given the training data set $\D$ consisting of the $N$ pilots in a  frame, frequentist learning produces an optimized single vector $\phi_\D^*$, while Bayesian learning returns a distribution $q^*(\phi|\D)$ on the model parameter space $\Phi$ \cite{blundell2015weight,simeone2022machine}. 
We denote as $p(y|x,\D)$ the resulting optimized predictive distribution which is either $p(y|x,\phi_{\D}^*)$ for frequentist learning, or the ensemble $\E_{\rvphi\sim q^*(\phi|\D)} [p(y|x,\rvphi)]$ for Bayesian learning.

\subsecnegspace
\subsection{Set Predictors}

A \emph{set predictor} is defined as a set-valued function $\Gamma(\cdot|\D): \mathcal{X} ~\rightarrow ~2^\mathcal{Y}$ that maps an input $x$ to a subset of the output domain $\mathcal{Y}$ based on data set $\D$. We denote the size of the set predictor for input $x$ as $|\Gamma(x|\D)|$. As illustrated in the example of Fig.~\ref{fig: fig_tikz_qpsk_prob_pred}, it depends in general on input $x$, and can be taken as a measure of the uncertainty of the set predictor.

The performance of a set predictor is evaluated in terms of coverage and inefficiency. \emph{Coverage} refers to the probability that the true label is included in the predicted set; while \emph{inefficiency} refers to the average size $|\Gamma(x|\D)|$ of the predicted set. There is clearly a trade-off between two metrics. 

Formally, the \emph{coverage} level of set predictor $\Gamma$ is the probability that the true output $y$ is included in the prediction set $\Gamma(x|\D)$ for a test pair $z=(x,y)$. This can be expressed as $\mathrm{coverage}(\Gamma) ~= \Prob~\big(\rv{y}\in \Gamma(\rv{x}|\mathbfcal{D})\big)$, where the probability $\Prob(\cdot)$ is taken over the ground-truth joint distribution of the involved random variables. When setting as target design a \emph{miscoverage level} $\alpha \in [0,1]$, the set predictor $\Gamma$ is said to be $1-\alpha$-\emph{valid} if
\begin{align}
    \mathrm{coverage}(\Gamma) = \Prob\big(\rv{y} \in \Gamma(\rv{x}|\mathbfcal{D})\big) \geq 1-\alpha.
    \label{eq: set validity}
\end{align}

It is straightforward to design a valid set predictors even for the restrictive case of miscoverage level $\alpha=0$ by producing the full set $\Gamma(x|\D)=\mathcal{Y}$ for all inputs $x$. One should, therefore, also consider the inefficiency of predictor $\Gamma$. The \emph{inefficiency} of set predictor $\Gamma$ is defined as the average predictive set size
\begin{align}
    \mathrm{inefficiency}(\Gamma) = \E \Big[ \big|\Gamma(\rv{x}|\mathbfcal{D})\big| \Big], \label{eq: ineff(Gamma) = E | Gamma |}
\end{align}
where the average is taken over the data set $\rvD$ and the test pair $(\rv{x},\rv{y})$ following their exchangeable joint distribution $p_0(\D,(x,y))$.

We note that the coverage condition \eqref{eq: set validity} is practically relevant if the learner produces multiple predictions using independent data set $\D$, and is tested on multiple pairs $(x,y)$. In fact, in this case, the probability in \eqref{eq: set validity} can be interpreted as the fraction of predictions for which the set predictor $\Gamma(x|\D)$ includes the correct output. This situation reflects well the setting of interest in which a different demodulator is designed for each frame.

\subsecnegspace
\subsection{\Naive Set Predictors from Probabilistic Predictors}\label{sec: Set Predictors from Probabilistic Predictors}

Given a probabilistic predictor $p(y|x,\D)$, one can construct a set predictor by relying on the confidence levels reported by the model. To this end, one can construct the smallest subset of the output domain $\mathcal{Y}$ that covers a fraction $1-\alpha$ of the probability designed by model $p(y|x,\D)$ given an input $x$. Given that probabilistic predictors are typically poorly calibrated, this approach generally does not satisfy condition \eqref{eq: set validity} for the given desired miscoverage level $\alpha$.

\secnegspace
\section{Conformal Prediction}\label{sec: Conformal Prediction}

\subsecnegspace
\subsection{Nonconformity Scores}\label{sec: Nonconformity Scores}

Conformal prediction relies on some form of validation to calibrate a \naive predictor. For any given test input $x$, a value $y^\prime\in\mathcal{Y}$ for input $x$ is included in the prediction set if $(x,y^\prime)$ ``conforms'' well with the validation data. To formalize CP, we define a \emph{nonconformity (NC) score} as a function that maps a pair $z=(x,y)$ and a data set $\D$ with $N$ samples to a real number, measuring how dissimilar the data point $z$ is to the data points in the fitting data set $\D$. An NC score must be invariant to permutations of the samples in the data set $\D$. 

Given a trained probabilistic model $p(y|x,\D)$, which may be frequentist or Bayesian, an NC score can be obtained as the log-loss
\begin{equation}
    \NC(z=(x,y)|\D) = -\log p(y|x,\D) \label{eq: NC classification}
\end{equation}
as long as the training algorithm used to derive the predictor $p(y|x,\D)$ is invariant to permutations of the data set $\D$. Note that \eqref{eq: NC classification} measures how poorly the sample $(x,y)$ conforms with respect to the data set $\D$ via the trained model $p(y|x,\D)$: 
If the sample $(x,y)$ is ``similar'' to the points in the set $\D$, the log-loss will tend to be small.

\subsecnegspace
\subsection{Validation-Based Set Predictors}\label{sec: Validation-Based Set Predictors}

\emph{Validation-based (VB)-CP} set predictors partition the available set $\D=\Dtr\cup\Dval$ into training set $\Dtr$ with $\Ntr$ samples and a validation set $\Dval$ with $\Nval=N-\Ntr$ samples.

Given a test input $x$, for each candidate output $y^\prime$ in $\mathcal{Y}$, the NC score $\NC((x,y^\prime)|\Dtr)$ is evaluated by using the training data $\Dtr$. The NC score $\NC((x,y^\prime)|\Dtr)$ is compared to the NC scores $\NC(z^\text{val}[i]|\Dtr)$ evaluated on all points $z^\text{val}[i], i=1,\dots,\Nval$ in the validation set $\Dval$. If the pair $(x,y^\prime)$ has a lower (or equal) NC score than a portion of at least $\lfloor \alpha(N^\text{val}+1)\rfloor/N^\text{val}$ of the validation NC scores, then the candidate label $y^\prime$ is included in the VB prediction set $\Gamma_{\alpha}^\text{VB}(x|\D)$. Accordingly, the VB-CP set predictor is obtained as
\begin{eqnarray}
    \Gamma_{\alpha}^\text{VB}(x|\D) =  \Big\{ y^\prime\in\mathcal{Y} \Big| \negspaceF && \NC((x,y^\prime)|\Dtr)  \label{eq: prediction set VB}\\ 
    \negspaceF&& \leq \quan_{1-\alpha} \big(\{ \NC(z^\text{val}[i]|\Dtr)\}_{i=1}^{N^\text{val}}\big) \Big\}. \nonumber
\end{eqnarray}
where the \emph{$(1-\alpha)$-empirical quantile} $\quan_{1-\alpha}\ \big(\{r[i]\}_{i=1}^{N}\big)$ for a set of $N$ real values $\{r[i]\}_{i=1}^{N}$ is the $\big\lceil  (1-\alpha)(N+1) \big\rceil$ th smallest value of the set $\{r[i]\}_{i=1}^{N}\cup\{+\infty\}$.

It is known from \cite{vovk2005algorithmic} that the VB-CP set predictor satisfies the coverage condition \eqref{eq: set validity}.
In terms of computational complexity, given $\Nte$ test inputs, predictor $p(y|x,\D)$ should be trained only once based on the training set $\Dtr$. This is followed by $\Nte |\mathcal{Y}| + N^\text{val}$ evaluations of the NC scores to obtain the $\Nte$ set predictions for all test points.

\subsecnegspace
\subsection{Cross-Validation-Based Set Predictors}\label{sec: Cross-Validation-Based Set Predictors}

VB-CP has the computational advantage of requiring a single training step, but the split into training and validation data causes the available data to be used in an inefficient way, while may in turn yield set prediction with large average size \eqref{eq: ineff(Gamma) = E | Gamma |}. Unlike VB-CP methods, cross-validation-based (CV) CP methods train multiple models, each using a subset of the available data set. As a result, CV-CP increases the computational complexity as compared to VB-CP, while generally reducing the inefficiency of set prediction \cite{barber2021predictive,romano2020classification}. 
Given a data set $\D=\{z[i]\}_{i=1}^N$ of $N$ points, the CV predictor fits $N$ models, one for each of the leave-one-out (LOO) sets $\big\{\D\setminus\{z[i]\}\big\}_{i=1}^N$ that exclude one of the points $z[i]$, which will play the role of validation \cite{barber2021predictive,romano2020classification}. Then, prediction on an input $x$ is done by evaluating the NC scores $\NC\big((x,y^\prime)\big|\D\setminus\{z[i]\}\big)$ of all prospective pairs $(x,y^\prime)$, using all available $N$ fitted models based on $N$ LOO sets $\D\setminus\{z[i]\}$, as well as the NC scores $\NC\big(z[i]\big|\D\setminus\{z[i]\}\big)$ for all validation data points. Accordingly, by including a candidate $y^\prime \in \mathcal{Y}$ if the NC score for $(x,y^\prime)$ is smaller (or equal) than a portion of at least $\lfloor \alpha(N^\text{val}+1)\rfloor/N^\text{val}$ of the validation data points, the CV-CP produces set predictor
\begin{eqnarray}
    \Gamma_{\alpha}^\text{CV}(x|\D) =  \bigg\{ y^\prime\in\mathcal{Y} \bigg| \negspaceF&&\negspaceF \sum_{i=1}^N \indicator \Big( \NC\big((x,y^\prime)\big|\D\setminus\{z[i]\}\big) \label{eq: prediction set CV classification} \\
    \negspaceF&&\negspaceF\negspaceF \leq \NC\big(z[i]\big|\D\setminus\{z[i]\}\big) \Big) \geq \lfloor\alpha(N+1)\rfloor \bigg\},\nonumber
\end{eqnarray}
with indicator function   $\indicator(\cdot)$ ($\indicator(\text{true})=1 $ and $\indicator(\text{false})=0$).

$K$-fold CV is a generalization of CV-CP set predictors that strike a balance between complexity and inefficiency by reducing the total number of model training phases to $K$ where $K\in\{2,\dots,N\}$ and $N/K$ is an integer. It then  trains $K$ models over leave-fold-out data sets, each of size $N-K$, and as validation uses the entire $N$ data set  \cite{barber2021predictive}.

By \cite[Theorms 1 and 4]{barber2021predictive}, CV-CP \eqref{eq: prediction set CV classification} satisfies the inequality
\begin{eqnarray}
    \Prob\big( \rv{y} \in \Gamma_{\alpha}^\text{CV}(\rv{x}|\rvD) \big) \geq 1-2\alpha \label{eq: CV is valid}
\end{eqnarray}
and its $K$-fold version $K$-CV satisfies the condition
\begin{equation}
    \Prob\big( \rv{y} \in \Gamma_{\alpha}^\text{$K$-CV}(\rv{x}|\rvD) \big) 
    \geq 1-2\alpha -\min\Big\{\tfrac{2(1-1/K)}{N/K+1},\tfrac{1-K/N}{K+1}\Big\}.
\end{equation}
Therefore, validity for both schemes is guaranteed for the larger miscoverage level of $2\alpha$. Accordingly, one can achieve miscoverage level of $\alpha$, satisfying \eqref{eq: set validity}, by considering the CV-CP set predictor 
with half of the target level $\alpha$. That said, numerical evidence reported in \cite{barber2021predictive} and \cite{romano2020classification} suggests that this is practically unnecessary.

\secnegspace
\section{Experiments and Conclusions}\label{sec: Experiments and Conclusions}

As in \cite{park2021fewpilots,cohen2021learning}, demodulation is implemented via a neural network model $p(y|x,\phi)$ consisting of a fully connected network with three hidden layers with ReLU activations, and softmax activation for the last layer. The amplitude and phase imbalance parameters in \eqref{eq: demodulation channel model}-\eqref{eq: iq imbalance as matrix} are independent and distributed as $\rvepsilon \sim \Betadist(\epsilon / 0.15 | 5,2)$ and $\rvdelta \sim \Betadist(\delta / 0.15^{\circ} | 5,2)$, respectively \cite{park2021fewpilots}. The SNR is set to 5 dB. The NC score \eqref{eq: NC classification} is evaluated as follows. For frequentist learning, the trained model $\phi_\D$ is obtained via $120$ gradient descent update steps for the minimization of the cross-entropy training loss with learning rate $0.2$. For Bayesian learning, we implemented stochastic gradient Langevin dynamics (SGLD) updates with burn-in period of $100$, ensemble size $20$, and learning rate $0.2$ \cite{welling2011bayesian}. We compare the \naive set predictor described in Sec.~\ref{sec: Set Predictors from Probabilistic Predictors}, 
which provides no formal coverage guarantees, with the CP set demodulation methods introduced in this work. We target the miscoverage level $\alpha=0.1$.

\begin{figure}[htb]
    \begin{minipage}[b]{1.0\linewidth}
        \centering
        \includegraphics[trim=0.0cm 0cm 1.0cm 0cm, clip, width=7.7cm]{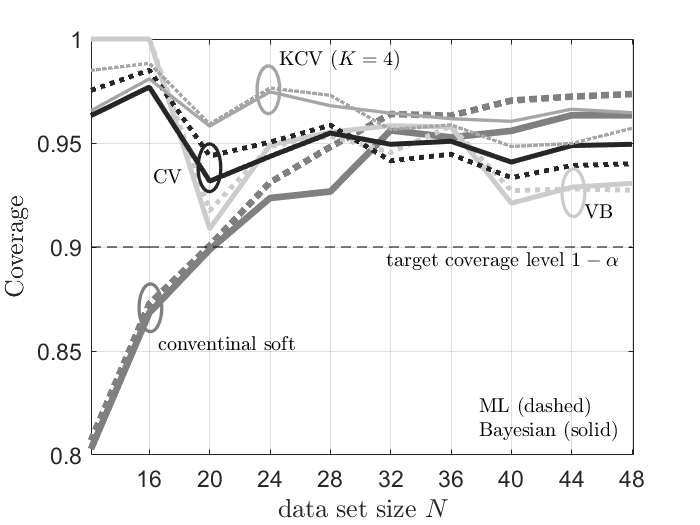}
        \vspace{-0.4cm}
        \caption{Coverage for \naive predictor, validation-based (VB) conformal predictor \eqref{eq: prediction set VB}, cross-validation-based (CV) conformal predictor, \eqref{eq: prediction set CV classification}, and the $K$-fold CV ($K$-CV) predictor as a function of the number of pilots $N$. The NC scores are evaluated either using frequentist learning (dashed lines) or Bayesian learning (solid lines).
        }
        \label{fig: coverage_vs_N_demod}
        \vspace{-0.3cm}
    \end{minipage}
\end{figure}
\begin{figure}[htb]
    \begin{minipage}[b]{1.0\linewidth}
        \centering
        \includegraphics[trim=0.0cm 0cm 1.0cm 0cm, clip, width=7.7cm]{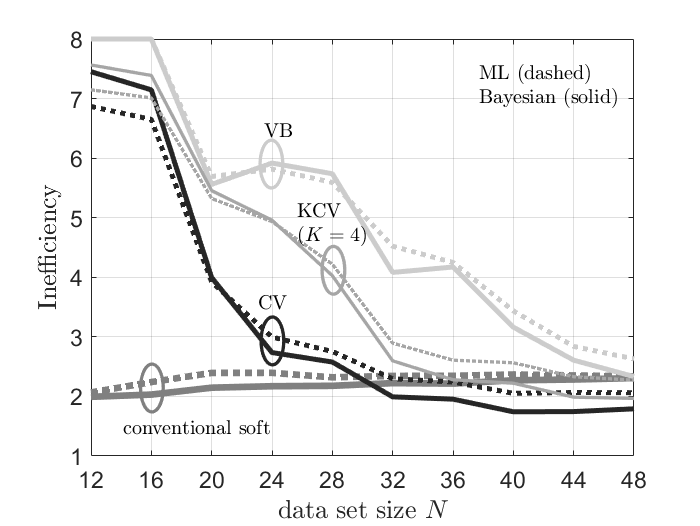}
        \vspace{-0.4cm}
        \caption{Average set prediction size (inefficiency) for the same setting of Fig.~\ref{fig: coverage_vs_N_demod}.}
        \label{fig: ineff_vs_N_demod}
        \vspace{-0.3cm}
    \end{minipage}
\end{figure}

Fig.~\ref{fig: coverage_vs_N_demod} shows the empirical coverage level and Fig.~\ref{fig: ineff_vs_N_demod} shows the empirical inefficiency, both evaluated on a test set with $100$ samples, as function of the size $N$ of the available data set $\D$. We further average the results for 50 independent frames, each corresponding to independent draws of pilot and data symbols from the ground truth distribution. From Fig.~\ref{fig: coverage_vs_N_demod}, we first observe that the \naive set predictor, with both frequentist and Bayesian learning, does not meet the desired coverage level in the regime of a small number $N$ of available samples. In contrast, confirming the theoretical guarantees presented in Sec.~\ref{sec: Conformal Prediction}, all CP methods provide coverage guarantees, achieving coverage rates above $1-\alpha$. From Fig.~\ref{fig: ineff_vs_N_demod}, we observe that the size of the prediction sets, and hence the inefficiency, decreases as the data set size, $N$, increases. Furthermore, due to their efficient use of the available data, CV and $K$-CV predictors have a lower inefficiency as compared to VB predictors. Finally, Bayesian NC scores are generally seen to yield set predictors with lower inefficiency, confirming the merits of Bayesian learning in terms of calibration. 

Overall, the experiments  confirm that all the CP-based predictors are all well-calibrated with small average set prediction size, unlike \naive set predictors that built directly on the self-reported confidence levels of conventional probabilistic predictors.

\vfill\pagebreak

\bibliographystyle{IEEEbib}
\bibliography{my_bib.bib} 

\end{document}